\documentclass[12pt]{article}

\usepackage{graphicx}
\usepackage{color}
\usepackage{epsfig}
\usepackage{latexsym}
\usepackage{bm}
\usepackage{ulem}
\usepackage{color}
\usepackage{dcolumn}
\usepackage{amsmath}
\usepackage{amssymb}
\usepackage{bm}
\usepackage{hyperref}
\usepackage{latexsym}
\usepackage{color}
\usepackage{subfigure}
\usepackage{breqn}
\usepackage{setspace}

\def\beq{\begin{equation}}
\def\eeq{\end{equation}}                         
\def\bea{\begin{eqnarray}}                           
\def\eea{\end{eqnarray}}

\begin{document}

\begin{center}
{\Large{\bf Ordering Kinetics in the Active Model B}} \\
\ \\
\ \\
by \\
Sudipta Pattanayak$^1$, Shradha Mishra$^2$ and Sanjay Puri$^3$ \\
$^1$S.N. Bose National Centre for Basic Sciences, JD Block, Sector III, Salt Lake City, Kolkata -- 700106, India. \\
$^2$Department of Physics, Indian Institute of Technology BHU, Varanasi -- 221005, India. \\
$^3$School of Physical Sciences, Jawaharlal Nehru University, New Delhi -- 110067, India. \\
\end{center}

\begin{abstract}
We undertake a detailed numerical study of the {\it Active Model B} proposed by Wittkowski et al. [Nature Comm. {\bf 5}, 4351 (2014)]. We find that the introduction of activity has a drastic effect on the ordering kinetics. First, the domain growth law shows a crossover from the usual Lifshitz-Slyozov growth law for phase separation ($L \sim t^{1/3}$, where $t$ is the time) to a novel growth law ($L \sim t^{1/4}$) at late times. Second, the correlation function exhibits dynamical scaling for a given activity strength $\lambda$, but the scaling function depends on $\lambda$.
\end{abstract}

\newpage

\section{Introduction}
\label{s1}

The ordering kinetics of an assembly of {\it self-propelled particles} (SPPs) is of great current interest \cite{mjr13,mc12}. Each SPP converts its internal energy into a systematic movement, thereby violating time-reversal-symmetry (TRS) \cite{jyl17}. These systems range in size from a few micrometers (e.g., actin and tubulin filaments, molecular motors \cite{nsm97,hnk87}, unicellular organisms such as amoeba and bacteria \cite{jb98}) to several meters (e.g., schools of fish \cite{jb98}, bird flocks \cite{cww19}, human crowds \cite{ph97}). Some SPPs, which are symmetric in shape but have a preferred direction of motion, are also called {\it active Brownian particles} (ABPs) \cite{fm12,pdk19,ms17}. Many lab-designed particles like {\it active colloids} \cite{hjr07,eh10,tsh11,vbv11,pss13} and {\it active Janus particles} \cite{csf17,pub18} are examples of ABPs. They have many technological applications like directional transport \cite{pdk19,ms17}, sorting of particles \cite{slr17,rr18}, etc.

One of the interesting features of a collection of ABPs is that they show {\it motility induced phase separation} (MIPS) at much lower densities than their passive analogs \cite{fm12,tc08,ct15,ct13}. As the name suggests, this interesting phenomenon is solely due to the motility of each individual agent. Many studies have focused on understanding MIPS \cite{ssc18}. A recent paper by Wittkowski et al. \cite{wts14} introduced a phenomenological model for MIPS, and used it to study domain growth kinetics in a collection of ABPs. Their dynamical update equation for the local order parameter of the system is similar to the {\it Cahn-Hilliard (CH) equation} \cite{ch58} or {\it Model B} \cite{hh77} for a conserved order parameter. The active nature of the ABPs is modeled via an additional term (of strength $\lambda$) which cannot be derived from an equilibrium free energy functional. The resultant update equation is termed the {\it Active Model B} (AMB) \cite{wts14}. Wittkowski et al. have also presented preliminary numerical studies of coarsening kinetics in the AMB. They made two important observations: \\
(a) The domain growth kinetics is not severely affected by the additional activity term. \\
(b) The static phase diagram is altered due to a pressure jump across interfaces.

In this paper, we undertake a detailed theoretical study of domain growth kinetics in the AMB with a critical composition. We will present results for an off-critical composition at a later stage. Our results are complementary to those of Wittkowski et al. However, there are important points of difference, which we highlight below.

We consider the evolution of the AMB from a homogeneous and disordered initial condition, as is customary in studies of domain growth \cite{pw09,ab94}. The evolution is characterized by the emergence and growth of domains or ABP clusters. The domain size grows as a power law in time $t$: $L(t) \sim t^{1/z}$, where $z$ is the dynamic growth exponent. In the absence of activity ($\lambda = 0$), it is well-known that $z=3$, which is referred to as the {\it Lifshitz-Slyozov} (LS) growth law \cite{pw09,ab94}. We find that the active term has a major impact on the coarsening kinetics. Most importantly, $z$ shows a crossover from $z = 3$ at early times to $z = 4$ at late times. The crossover time $t_{c}$ decreases as a power law with the activity strength, $t_c \sim \lambda^{-3/2}$. Moreover, the density correlation function shows dynamical scaling for a given $\lambda$, but the scaling function varies with $\lambda$.

This paper is organized as follows. In Sec.~\ref{s2}, we introduce the AMB proposed by Wittkowski et al. In Sec.~\ref{s3}, we present detailed numerical results for domain growth in the AMB. In Sec.~\ref{s4}, we summarize our main conclusions.

\section{Model}
\label{s2}

We consider an assembly of ABPs on a two-dimensional ($d=2$) substrate and study the dynamics of the system. The ABPs move with a self-propulsion speed $v_0$, and their characteristic rotation frequency is $\tau^{-1}$. The local density of the ABPs is denoted as $n(\vec{r},t)$, where $\vec{r}$ is the position vector. The values of $n$ lie in the interval $[0,1]$. The corresponding order parameter is $\psi(\vec{r},t) = 2 n(\vec{r},t) - 1$, so that regions with $\psi > 0$ are enriched in particles. This system can be modeled by a coarse-grained partial differential equation for $\psi(\vec{r},t)$. The derivation of the hydrodynamic equation can be found in Ref.~\cite{ct15}. The resultant model can be expressed as a continuity equation for the conserved order parameter:
\bea
\frac{\partial}{\partial t} \psi(\vec{r},t) &=& -\vec{\nabla} \cdot \vec{J}(\vec{r},t) , \nonumber \\
\vec{J}(\vec{r},t) &=& - \vec{\nabla} \mu (\vec{r},t) ,
\label{eqn1}
\eea
where $\vec{J}(\vec{r},t)$ is the current. The corresponding chemical potential is \cite{wts14}
\beq
\mu (\vec{r},t) = -\psi(\vec{r},t) + \psi(\vec{r},t)^{3} - \nabla^{2}\psi(\vec{r},t) + \lambda|\vec{\nabla} \psi(\vec{r},t)|^{2} .
\label{eqn3}
\eeq

The above equations are formulated in dimensionless units. These are obtained by rescaling space by a persistence length $v_{0} \tau$, and time by the relaxation time $\tau$. The chemical potential $\mu$ in Eq.~(\ref{eqn3}) is the sum of bulk ($\mu_{0}$) and gradient ($\mu_{1}$) contributions. The bulk part is the same as for Model B, $\mu_{0} = -\psi({r},t) + \psi(\vec{r},t)^{3}$, and is derived from the bulk free-energy density of a symmetric Ginzburg-Landau (GL) $\psi^4$-field-theory \cite{pw09}. The gradient term can be written as the sum of two terms, $\mu_{1} = \mu_{1}^{p} + \mu_{1}^{a}$. Here, $\mu_{1}^{p} = -\nabla^2 \psi$ is derivable from the square-gradient term in the GL free-energy density. The term $\mu_{1}^{a}$ is the {\it active term} which breaks TRS and has strength $\lambda$, which is a tunable parameter in our study. This term is not obtainable as the derivative of a free energy, and its origin is similar to the lowest-order nonlinear interfacial  diffusion term in the Kardar-Parisi-Zhang or KPZ equation \cite{kpz86}. Therefore, the proposed update equation for $\psi(\vec{r}, t)$ in the AMB is
\begin{equation}
\frac{\partial}{\partial t} \psi(\vec{r},t) = \vec{\nabla} \cdot \left[ \vec{\nabla} (-\psi + \psi^{3} - \nabla^{2}\psi + \lambda |\vec{\nabla} \psi|^{2}) \right] . 
\label{eqn_activemodelB}
\end{equation}

\begin{figure}[ht]
\centering
\includegraphics[width=0.77\linewidth]{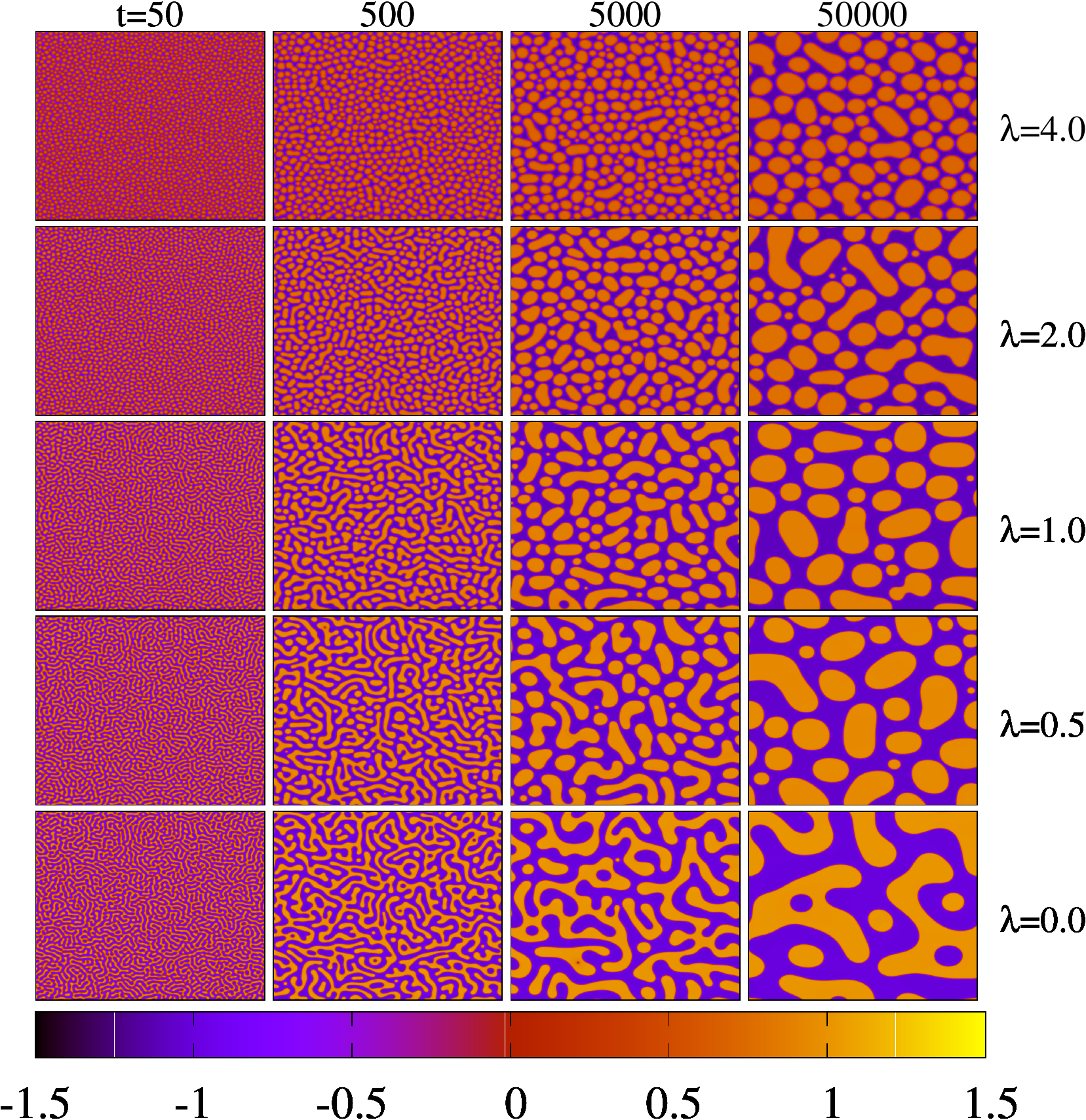}
\caption{(Color online) Evolution snapshots of the system at different times $(t)$ for various values of activity $(\lambda)$. The system size is $L^2 = 512^2$, and the average order parameter is $\psi_0 = 0$, corresponding to a critical composition. The color bar denotes the range of $\psi$-values.}
\label{f1}
\end{figure}

We numerically integrated Eq.~(\ref{eqn_activemodelB}) on a $d=2$ lattice of size $N^2$ with periodic boundary conditions. We used a simple Euler discretization scheme with mesh sizes $\Delta x = 1.0$ and $\Delta t = 0.01$. The space mesh is small enough to resolve coarsening interfaces, and the time mesh is adequate to ensure stability of the numerical scheme. The initial condition for a run consisted of the order parameter field $\psi (\vec{r},0)$ having small-amplitude fluctuations about an average value $\psi_0 = 0$. This case corresponds to a {\it critical} composition with an equal number of occupied and empty sites. All cases with $\psi_0 \neq 0$ are referred to as {\it off-critical} systems. In our simulations, $\lambda$ is varied from $0$ to $4$. The latter is the largest value of $\lambda$ for which we can have coexisting domains \cite{wts14}. For $\lambda=0$, the model reduces to the standard Model B \cite{pw09}. All results presented here are for lattice sizes $N=512$, and statistical quantities are averaged over 100 independent runs.

\section{Results}
\label{s3}

First, we study the evolving domain morphology for different values of $\lambda$. The snapshots of the local order parameter $\psi(\vec{r},t)$ for $\lambda=0.0,0.5,1.0,2.0,4.0$ at different times are shown in Fig.~\ref{f1}. The light and dark regions represent $\psi > 0$ and $\psi < 0$, respectively. There is an asymmetry between the saturation (fixed point) values of $\psi$ for the two regions. The static kink solution $\psi_s(z)$ of Eq.~(\ref{eqn_activemodelB}) is obtained from \cite{wts14}
\begin{equation}
-\psi_s + \psi_s^3 - \frac{d^2 \psi_s}{d z^2} + \lambda \left( \frac{d\psi_s}{dz} \right)^2 = \mu_s ,
\label{kink}
\end{equation}
where the static chemical potential $\mu_s$ is non-zero for $\lambda \neq 0$. For small values of $\lambda$, a perturbative calculation yields
\begin{equation}
\mu_s(\lambda) = \frac{4}{15} \lambda + O(\lambda^3) .
\end{equation}
It is clear from Eq.~(\ref{kink}) that $\mu_s(-\lambda) = -\mu_s(\lambda)$, so that there are only odd terms in the expansion of $\mu_s(\lambda)$. At large distances ($z \rightarrow \pm \infty$) from the kink center, we can ignore the derivative terms in Eq.~(\ref{kink}). Thus, the saturation values of the order parameter are obtained from
\begin{equation}
-\psi_s + \psi_s^3 = \mu_s(\lambda) ,
\end{equation}
i.e., magnetization of a ferromagnet in a magnetic field $\mu_s$. We can obtain the exact solutions ($\psi_0, \psi_1, \psi_2$) of this cubic equation. The corresponding perturbative result is
\begin{eqnarray}
\psi_1 &=& +1 + \frac{\mu_s}{2} , \nonumber \\
\psi_0 &=& -\frac{\mu_s}{2} , \nonumber \\
\psi_2 &=& -1 + \frac{\mu_s}{2} .
\end{eqnarray}

Clearly, the magnitude of $\psi_s$ in the light phase ($\psi_1$) is higher than that in the dark phase ($\psi_2$) for $\lambda > 0$. Therefore, the fraction of the light phase ($\phi_1$) is less than that of the dark phase ($\phi_2$): $\phi_2 - \phi_1 = 2\lambda/15 + O(\lambda^3)$. Hence, we see a droplet morphology at late times rather than the bicontinuous morphology which is characteristic of Model B, seen in the bottom row. We note that Eq.~(\ref{eqn_activemodelB}) is symmetric under the transformation $\psi \rightarrow -\psi$ and $\lambda \rightarrow -\lambda$. Thus, the evolution morphologies for $\lambda < 0$ are analogous to those in Fig.~\ref{f1}, except the droplets would be of the particle-poor phase.

The domain morphology is quantitatively characterized by the two-point correlation function:
\beq
C(r,t)=\langle \psi(\vec{R} + \vec{r}, t) \psi(\vec{R}, t) \rangle - \langle \psi(\vec{R} + \vec{r}, t) \rangle \langle \psi(\vec{R}, t) \rangle,
\eeq
where the $\langle .. \rangle$ denotes an average over reference positions $\vec{R}$, spherical averaging over different directions, and 100 independent runs. It is apparent from Fig.~\ref{f1} that the evolution morphology is characterized by a single length scale $L(t)$. This results in the dynamical scaling of the correlation function \cite{pw09}:
\begin{equation}
C(r,t) = f(r/L) ,
\label{dysc}
\end{equation}
where $f(x)$ is the scaling function. The domain size $L(t)$ is defined as the characteristic scale over which the correlation function $C(r,t)$ decays to (say) $0.5$ times its maximum value at $r=0$.

\begin{figure}[ht]
\centering
\includegraphics[width=0.8\linewidth]{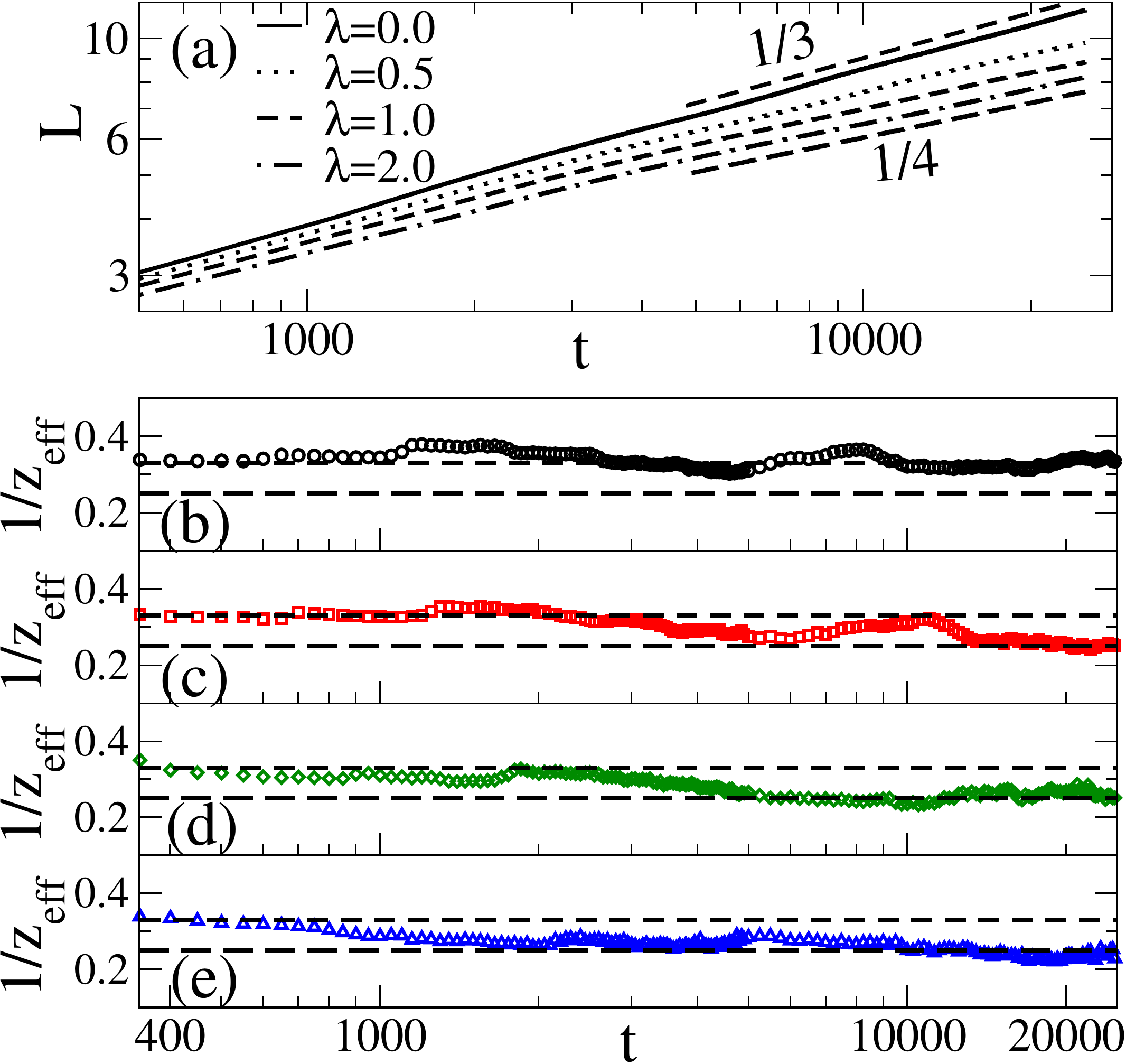}
\caption{(Color online) (a) Log-log plot of $L$ vs. $t$ for $\lambda = 0.0,0.5,1.0,2.0$. The lines labeled 1/3 and 1/4 denote the growth laws $L \sim t^{1/3}$ and $L \sim t^{1/4}$, respectively. (b)-(e) Plot of effective exponent ($1/z_{\rm eff}$) vs. $t$ for $\lambda = 0.0,0.5,1.0, 2.0$ (from top to bottom). In each frame, we have drawn horizontal lines at $1/z_{\rm eff} = 1/3$ and $1/z_{\rm eff} = 1/4$.}
\label{f2}
\end{figure}

\begin{figure}[ht]
\centering
\includegraphics[width=0.55\linewidth]{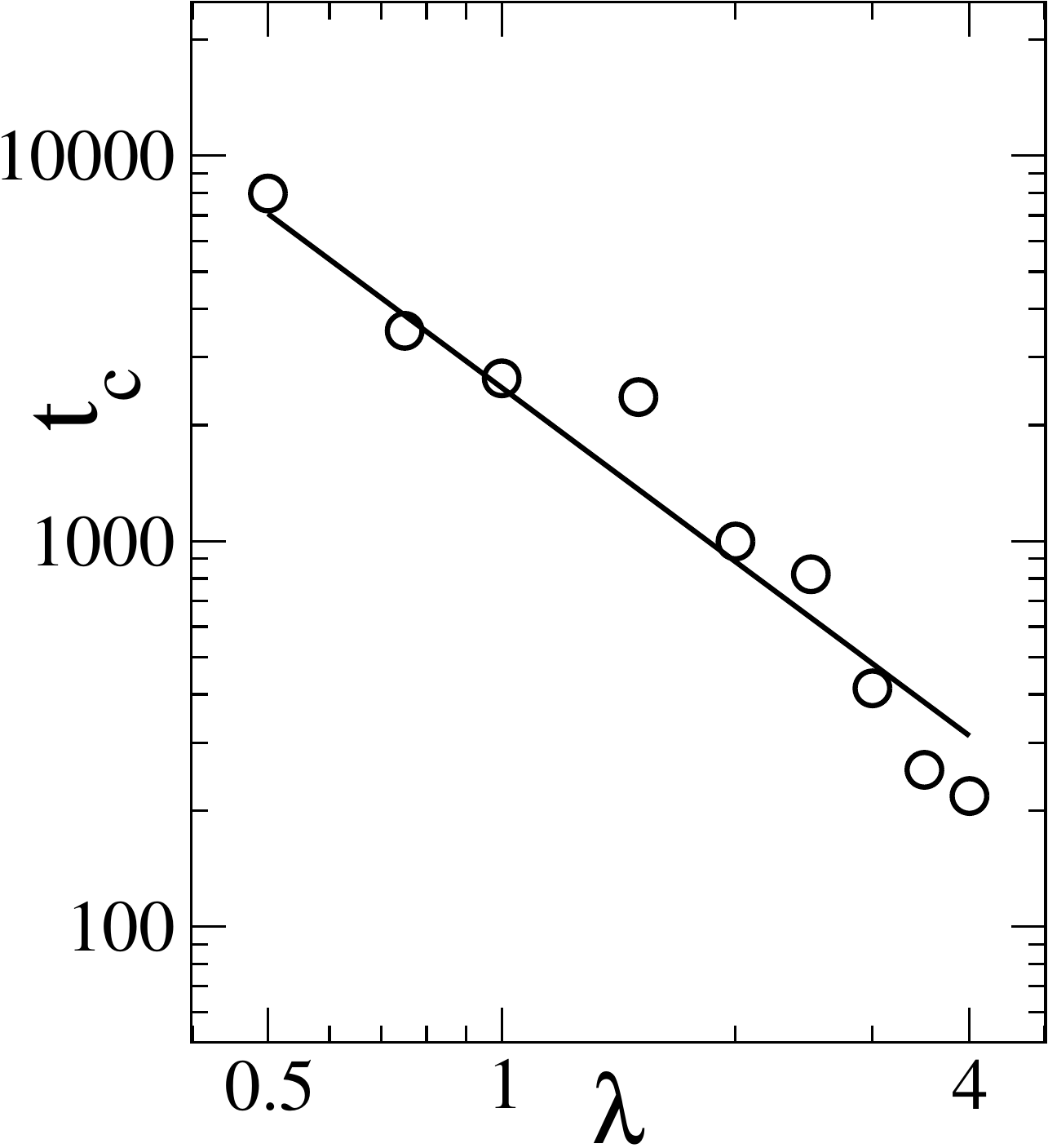}
\caption{Log-log plot of $t_{c}$ vs. $\lambda$. The solid line has a slope of $-3/2$.}
\label{f3}
\end{figure}

We plot $L(t)$ vs. $t$ on a log-log scale for $\lambda = 0.0,0.5,1.0,2.0$ in Fig.~\ref{f2}(a). We note that $z=3$ for Model B \cite{pw09}, whereas it crosses over from 3 to 4 for the AMB. To confirm this, we estimate the effective growth exponent $1/z_{\rm eff}$ as a function of $t$:
\begin{equation}
\dfrac{1}{z_{\rm eff}}=\dfrac{d \ln L}{d \ln t} .
\label{eqnexpo}
\end{equation}
In Fig.~\ref{f2}(b)-(e), we show the variation of $1/z_{\rm eff}$ with $t$ for the data sets in Fig.~\ref{f2}(a). These plots show considerable fluctuations, as is usual for the effective exponent. However, the overall trend is quite clear. For Model B ($\lambda=0.0$) in Fig.~\ref{f2}(b), $1/z_{\rm eff}$ fluctuates around $0.33$. For non-zero $\lambda$, $1/z_{\rm eff} \simeq 0.33$ at early times, and shows a crossover to $1/z_{\rm eff} \simeq 0.25$ at late times. The crossover occurs earlier for larger values of $\lambda$. We define the crossover time as $t_{c}=(t_1+t_2)/2$, where $t_1$ and $t_2$ are the times when $1/z_{\rm eff}$ crosses the value $0.29$ for the first and last times, respectively \cite{tc}. Fig.~\ref{f3} shows the variation of $t_{c}$ with $\lambda$ -- we see that $t_c$ decays algebraically with activity, $t_c \sim \lambda^{-\theta}$. The decay exponent is consistent with $\theta \simeq 1.5$. We can combine the early- and late-time behaviors of the length scale in a crossover function:
\bea
L(t) &=& t^{1/3} g(t/t_c) , \nonumber \\
g(x) &\rightarrow& a, \quad x \rightarrow 0 , \nonumber \\
g(x) &\rightarrow& b~x^{-1/12}, \quad x \rightarrow \infty ,
\eea
where $a$ and $b$ are constants. A domain growth law of the form $L \sim t^{1/4}$ has been proposed in the context of systems where phase separation occurs via surface diffusion rather than bulk diffusion \cite{pbl97,gbp05}. However, surface diffusion is clearly not the dominant mechanism for the crossover seen in the AMB, given the non-connected droplet morphology in Fig.~\ref{f1}.

To gain some insights on this crossover, let us examine the shrinking dynamics of a bubble of phase 2 ($\psi_2 < 0$) in a background of phase 1 ($\psi_1 > 0$). For simplicity, we consider a bubble in $d=3$ though our simulations correspond to $d=2$. The growth laws are expected to be the same for $d \geq 2$. The initial size of the bubble is $R(t=0)=R_0$, and the time-dependent size is $R(t)$. There is a direct relationship between the dynamics of $R(t)$ and $L(t)$. In the absence of activity ($\lambda = 0$), the shrinking of a $d=3$ bubble obeys \cite{ab94}
\begin{equation}
\frac{dR}{dt} = -\frac{\sigma}{2 R^2} ,
\label{bubble}
\end{equation}
where $\sigma$ is the surface tension. The solution of Eq.~(\ref{bubble}) is
\begin{equation}
R_0^3 - R(t)^3 = \frac{3 \sigma t}{2} .
\end{equation}
Thus, a bubble of size $R_0$ evaporates on the time-scale $t_0$, where $R_0 = (3 \sigma t_0/2)^{1/3}$. For the coarsening process, this means that all structures of size $\leq (\sigma t)^{1/3}$ have vanished by time $t$. This directly yields the well-known LS domain growth law, $L(t) \sim (\sigma t)^{1/3}$.

How is Eq.~(\ref{bubble}) modified by the activity $\lambda$? Given the invariance of Eq.~(\ref{eqn_activemodelB}) under the transformation $\lambda \rightarrow -\lambda$, $\psi \rightarrow -\psi$, we naively expect that there will be no odd powers of $\lambda$ in the generalization of Eq.~(\ref{bubble}). Let us confirm the validity of this to $O(\lambda)$, where the calculation is relatively straightforward. Our discussion follows that of Bray \cite{ab94}. The starting point of our discussion is Eq.~(\ref{eqn_activemodelB}) in more general notation:
\begin{equation}
\frac{\partial \psi}{\partial t} = \nabla^2 \left[ V'(\psi) - \nabla^2 \psi + \lambda |\vec{\nabla}\psi|^2 \right] \equiv \nabla^2 \mu ,
\label{amb}
\end{equation}
where we have introduced the $\psi^4$-potential, $V(\psi) = -\psi^2/2 + \psi^4/4$. In the late stages of domain growth, the bulk domains are saturated to their equilibrium values $\psi_1$ and $\psi_2$. There are small fluctuations about these values, e.g., $\psi = \psi_{1,2} + \phi$, which drive domain growth. We linearize Eq.~(\ref{amb}) in $\phi$ to obtain
\begin{equation}
\frac{\partial \phi}{\partial t} \simeq V''(\psi_{1,2}) \nabla^2 \phi - \nabla^4 \phi .
\label{amblin}
\end{equation}
As argued by Bray, we can neglect the $\partial/\partial t$ and $\nabla^4$ terms in Eq.~(\ref{amblin}). Therefore, in the bulk domains, the fluctuations obey the Laplace equation $\nabla^2 \phi = 0$. The linearized chemical potential in the bulk is $\mu \simeq V''(\psi_{1,2}) \phi - \nabla^2 \phi$. Again, the $\nabla^2 \phi$ term is negligible because the additional derivatives give extra powers of $L$ in the denominator. Thus, the chemical potential also obeys the Laplace equation in the bulk:
\begin{equation}
\nabla^2 \mu = 0 .
\end{equation}

Next, we derive the boundary conditions imposed on $\mu$ by the interfaces. We assume that the interfaces are almost equilibrated to the kink profile $\psi_s (z)$. We introduce the coordinate system $(g,\vec{s})$, where $g$ is normal to the interface (going from $\psi_2$ to $\psi_1$ with the interface located at $g=0$). The coordinates $\vec{s}$ are tangential to the interface. To obtain $\mu$ at the interfaces, we need
\begin{eqnarray}
\vec{\nabla} \psi &=& \frac{\partial \psi}{\partial g} \bigg|_{\vec{s}} \hat{g} , \nonumber \\
\nabla^2 \psi &=& \frac{\partial^2 \psi}{\partial g^2} \bigg|_{\vec{s}} + \frac{\partial \psi}{\partial g} \bigg|_{\vec{s}} \vec{\nabla} \cdot \hat{g} 
\equiv \frac{\partial^2 \psi}{\partial g^2} \bigg|_{\vec{s}} + \frac{\partial \psi}{\partial g} \bigg|_{\vec{s}} K .
\label{int}
\end{eqnarray}
In Eq.~(\ref{int}), $\hat{g}$ is the unit vector normal to the interface and $K$ is the local curvature. Thus, near the interface,
\begin{equation}
\mu = V'(\psi) -  \frac{\partial \psi}{\partial g} \bigg|_{\vec{s}} K - \frac{\partial^2 \psi}{\partial g^2} \bigg|_{\vec{s}} + \lambda \left(\frac{\partial \psi}{\partial g} \bigg|_{\vec{s}}\right)^2 .
\label{muint}
\end{equation}
We multiply Eq.~(\ref{muint}) by $\frac{\partial \psi}{\partial g} \big|_{\vec{s}}$, which is sharply peaked at $g=0$, and integrate $g$ from $-\infty$ to $\infty$. This yields
\begin{equation}
\mu \Delta \psi = \Delta V -  \alpha K - \int_{-\infty}^{\infty} dg~\frac{1}{2} \frac{\partial}{\partial g} \left(\frac{\partial \psi}{\partial g} \bigg|_{\vec{s}}\right)^2 + \beta \lambda ,
\label{mupsi}
\end{equation}
where
\begin{eqnarray}
\Delta \psi &=& \psi_1 - \psi_2 = 2+O(\lambda^2) , \nonumber \\
\Delta V &=& V(\psi_1) - V(\psi_2) = O(\lambda^2) , \nonumber \\
\alpha &=& \int_{-\infty}^{\infty} dg~\left(\frac{\partial \psi}{\partial g} \bigg|_{\vec{s}}\right)^2 , \nonumber \\
\beta &=& \int_{-\infty}^{\infty} dg~\left(\frac{\partial \psi}{\partial g} \bigg|_{\vec{s}}\right)^3 .
\end{eqnarray}

To obtain $\alpha$ and $\beta$, we need the static kink solution from Eq.~(\ref{kink}). We use the expansion
\begin{equation}
\psi_s (z) = \tanh \left(\frac{z}{\sqrt{2}} \right) + \lambda \psi_1(z) + O(\lambda^2) ,
\end{equation}
where the leading term is the kink solution for Model B. The resultant equation for $\psi_1(z)$ is
\begin{equation}
\frac{d^2 \psi_1}{d z^2} + \left[ 1 - 3 \tanh^2\left(\frac{z}{\sqrt{2}}\right)\right] \psi_1 = \frac{1}{2} \mbox{sech}^4 \left(\frac{z}{\sqrt{2}}\right) - \frac{4}{15} .
\label{psi1}
\end{equation}
From Eq.~(\ref{psi1}), it is simple to confirm that $\psi_1(-z) = \psi_1(z)$. Thus
\begin{eqnarray}
\alpha &=& \int_{-\infty}^{\infty} dg~\left[ \frac{1}{\sqrt{2}}~{\mbox{sech}}^2 \left( \frac{g}{\sqrt{2}} \right) + \lambda \frac{d\psi_1}{dg} \right]^2 \nonumber \\
&=& \frac{1}{2} \int_{-\infty}^{\infty} dg~{\mbox{sech}}^4 \left( \frac{g}{\sqrt{2}} \right) + O(\lambda^2) \nonumber \\
&=& \sigma + O(\lambda^2) ,
\end{eqnarray}
where $\sigma = 2\sqrt{2}/3$ is the surface tension for Model B. For $\beta$, we multiply Eq.~(\ref{kink}) by $d\psi_s/dz$ and integrate $z$ from $-\infty$ to $\infty$. This yields $\Delta V + \lambda \beta = \mu_s \Delta \psi$. Replacing this in Eq.~(\ref{muint}), we obtain the chemical potential on the interfaces as
\begin{equation}
\mu = -\frac{\alpha}{\Delta \psi} K + \mu_s .
\end{equation}

Recall that we are interested in the shrinking dynamics of a spherical bubble of radius $R(t)$. The center of the bubble is located at $r=0$. The chemical potential in the bulk is obtained as the solution of Laplace's equation $\nabla^2 \mu = 0$. The curvature for a $d=3$ bubble is $K=2/R$. Thus, the appropriate expression for $\mu (r)$ is
\begin{eqnarray}
\mu (r) &=& -\frac{2\alpha}{\Delta \psi} \frac{1}{R} + \mu_s , \quad r \leq R , \nonumber \\
&=& -\frac{2\alpha}{\Delta \psi} \frac{1}{r} + \mu_s , \quad r > R .
\label{mur}
\end{eqnarray}
In Eq.~(\ref{mur}),
\begin{equation}
\frac{2\alpha}{\Delta \psi} = \frac{\sigma + O(\lambda^2)}{1 + O(\lambda^2)} .
\end{equation}
The shrinking of the bubble obeys
\begin{eqnarray}
\frac{dR}{dt} &=& - \frac{1}{\Delta \psi} \vec{\nabla} \mu \big|_{r=R} \nonumber \\
&=& - \frac{2\alpha}{(\Delta \psi)^2} \frac{1}{R^2} \simeq - \frac{\sigma + O(\lambda^2)}{2} \frac{1}{R^2} ,
\end{eqnarray}
yielding the LS growth law with a $\lambda^2$-dependent prefactor.

\begin{figure}[ht]
\centering
\includegraphics[width=0.4\linewidth]{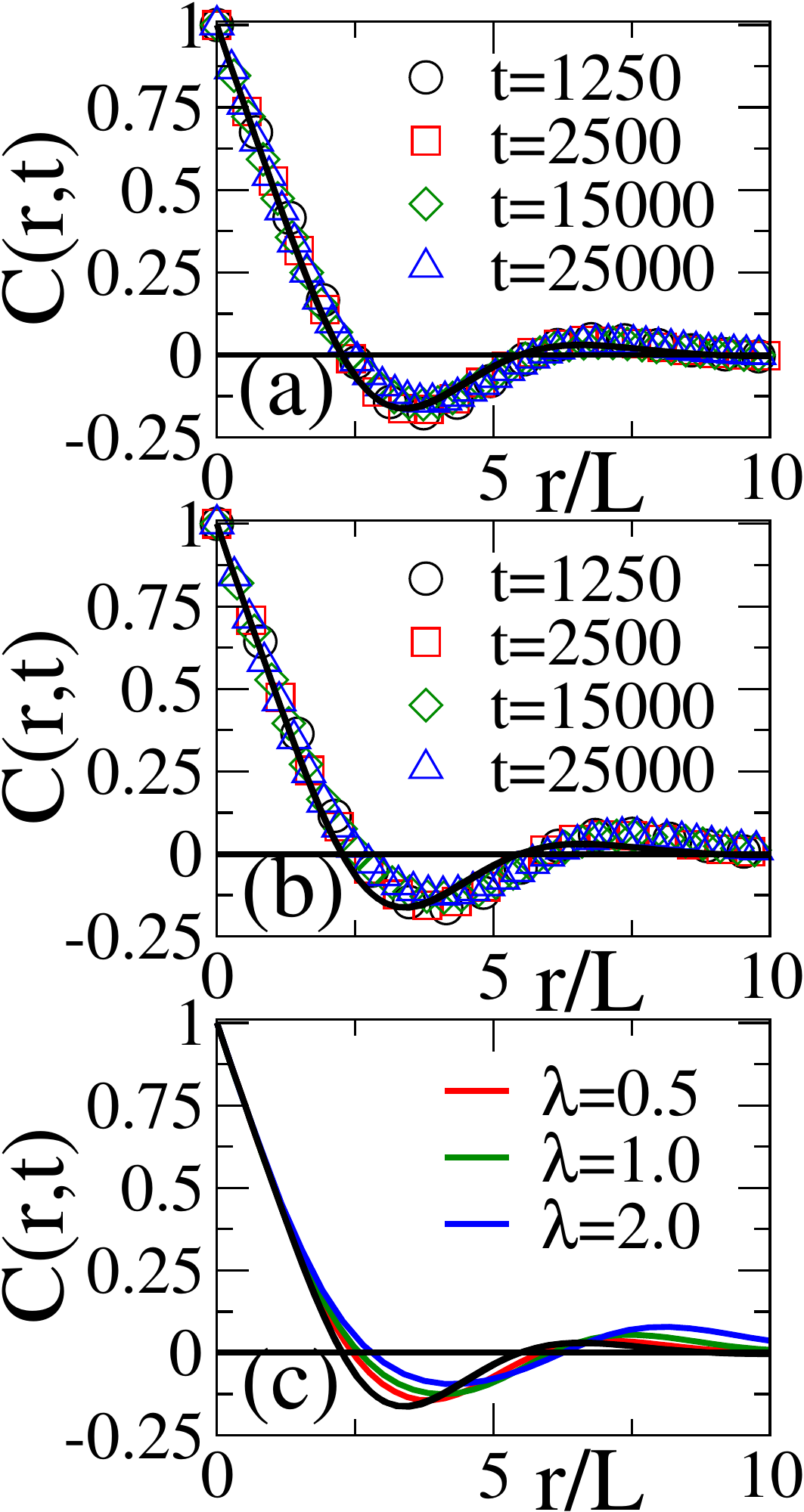}
\caption{(Color online) (a) Scaling plot of $C(r,t)$ vs. $r/L$ for $\lambda=0.5$. We plot data at the indicated times. The solid line denotes the scaling function for Model B at critical composition. (b) Analogous to (a), but for $\lambda=1.0$. (c) Scaling functions for different values of $\lambda$ at $t=25000$.}
\label{f4}
\end{figure}

From the above discussion, it is clear that a simple generalization of bubble dynamics to the case of non-zero background chemical potential is not adequate to capture the crossover in the growth law seen in the AMB. We need a more sophisticated theory, which accounts for the modification of $\alpha$ and $\beta$ in Eq.~(\ref{mupsi}) by the local curvature of the interface. This is beyond the scope of the present paper. of course, based on the numerical results, we can propose several simple phenomenological equations to capture the crossover in the growth law, e.g.,
\begin{equation}
\frac{dL}{dt} = \frac{a_1}{L^2 (1 + b_1 L)} ,
\end{equation}
or
\begin{equation}
\frac{dL}{dt} = \frac{a_2}{L^2 \sqrt{1 + b_2 L^2}} .
\end{equation}
However, such a venture would be speculative so we do not pursue it further.

Finally, let us study the dynamical scaling behavior of the correlation function $C(r,t)$ for different values of $\lambda$. In Fig.~\ref{f4}(a)-(b), we plot $C(r,t)$ vs. $r/L$ [see Eq.~(\ref{dysc})] for $\lambda=0.5$ and $1.0$, respectively. We superpose data at four different times, two at $t<t_c$, and two at $t>t_c$. The data sets neatly collapse and confirm scaling in the system. Clearly, the crossover in the growth law does not affect the system morphology. The solid curve in Fig.~\ref{f4}(a)-(b) denotes the scaling function for Model B. This is seen to differ from the AMB scaling functions. Recall that Model B shows a bicontinuous morphology, whereas the AMB shows a droplet morphology -- see Fig.~\ref{f1}. A droplet morphology would arise in Model B for an off-critical composition, with unequal fractions of the two species. In Fig.~\ref{f4}(c), we check whether the droplet scaling function is dependent on $\lambda$ by plotting $C(r,t)$ vs. $r/L$ for $\lambda=0.5,1.0,2.0$ at $t=25000 \gg t_c(\lambda)$. The scaling function changes continuously with $\lambda$. With increase in $\lambda$, the composition asymmetry of the two phases becomes more pronounced. This changes the relative fraction or effective ``off-criticality'' of the particle-rich and particle-poor phases. It is known that the scaling function depends on the off-criticality \cite{sp88}, which is reflected in Fig.~\ref{f4}(c).

\section{Summary}
\label{s4}

Let us conclude this paper with a summary and discussion of our results. We have undertaken a detailed numerical study of ordering kinetics in the {\it Active Model B} (AMB) proposed by Wittkowski et al. \cite{wts14}. The AMB consists of the usual Cahn-Hilliard (CH) equation or Model B for the kinetics of phase separation, supplemented by an active term of strength $\lambda$. The active term breaks time-reversal symmetry and is not derivable from a Ginzburg-Landau free energy. The CH equation exhibits static kink solutions connecting the particle-rich ($\psi=+1$) and particle-poor ($\psi=-1$) phases. These kink solutions correspond to a uniformly zero value of the chemical potential. In the AMB, there also exist static kink solutions for $\lambda \leq 4$. However, these differ from the CH kinks in two important ways: \\
(a) the composition of the two phases is asymmetric, \\
(b) the background chemical potential is non-zero.

Our numerical study of the AMB demonstrates two important features: \\
(a) The domain growth law crosses over from $L \sim t^{1/3}$ (the usual Lifshitz-Slyozov or LS law for phase separation) at early times to $L \sim t^{1/4}$ at late times. The crossover time-scale is consistent with the scaling behavior $t_c \sim \lambda^{-3/2}$. To gain some insights on this crossover, we have examined possible extensions of the dynamical equation for shrinking of a bubble. The obvious generalization to the case of non-zero background chemical potential is not adequate to account for the crossover. This only modifies the surface tension prefactor in the LS growth law by a $\lambda^2$-dependent factor. We need a more sophisticated modification which accounts for the manner in which local curvature is affected by the active term. \\
(b) The composition asymmetry of the two phases yields a droplet morphology, even when the overall composition is critical. This has important consequences for the dynamical scaling of the correlation function. For a given activity strength $\lambda$, the correlation function obeys dynamical scaling. Thus, the morphology is not affected by the crossover in the growth law. However, the scaling function has a continuous dependence on $\lambda$.

This problem is of great topical interest, given the current focus on active matter. We hope that our numerical results will motivate further analytical and numerical interest in this problem. Clearly, the outstanding problem is the development of an interface-kinetics theory to explain the crossover in the domain growth law.

\subsubsection*{Acknowledgments}

Sudipta Pattanayak would like to thank the Department of Physics, Indian Institute of Technology (BHU), Varanasi for their hospitality. Shradha Misra would like to thank the S.N. Bose National Centre for Basic Sciences, Kolkata for hosting her.


\begin{thebibliography}{99}
\bibitem{tc} We also define the crossover time in two other ways, (i) when $1/z_{\rm eff}$ crosses the value 0.33 for the last time; and (ii) when $1/z_{\rm eff}$ crosses the value 0.25 for the first time. We note a similar power-law variation for these definitions also.
\end{thebibliography}
\end{document}